# Is there a robust effect of mainland mutualism rates on species richness of oceanic islands?

Maximilian Pichler* & Florian Hartig

Theoretical Ecology, University of Regensburg, Regensburg, Germany

*corresponding author, maximilian.pichler@biologie.uni-regensburg.de

# Main

In island biogeography, it is widely accepted that species richness on island depends on the area and isolation of the island as well as the species pool on the mainland [1]. Delavaux et al.[2] suggest that species richness on oceanic islands also depends on the proportion of mutualists on the mainland, based on the idea that mutualists require specific interaction partners for their survival and thus have lower chances of establishment after successful immigration. As the proportion of mutualists increases towards the tropics, this effect could explain a weaker latitudinal diversity gradient (LDG) for oceanic islands. However, after re-analyzing their data, we have doubts if these conclusions are supported by the available data.

The first issue is that mutualism filter strength (the proportion of mutualists on the mainland) is only available at a limited set of sites. To interpolate it to the relevant points for the islands, Delavaux et al.[2] used predictions from a poorly fitting generalized linear additive model (GAM) that included only latitude as a predictor ($R^2$ averaged across three mutualism types = 0.141). This model fails to reflect systematic longitudinal patterns in mutualisms strength. It is well known that using covariates with large errors can lead to issues in regression models[3,5], in particular under predictor collinearity[4]. Interpolating mutualism strength with a random forest model based on latitude and longitude leads to substantially improved predictions (average $R^2$ = 0.389). When using the mutualism covariate predicted from this improved model in the main analysis, results are changed, and the effect of mutualism strength becomes non-significant and small (Fig. 1D).

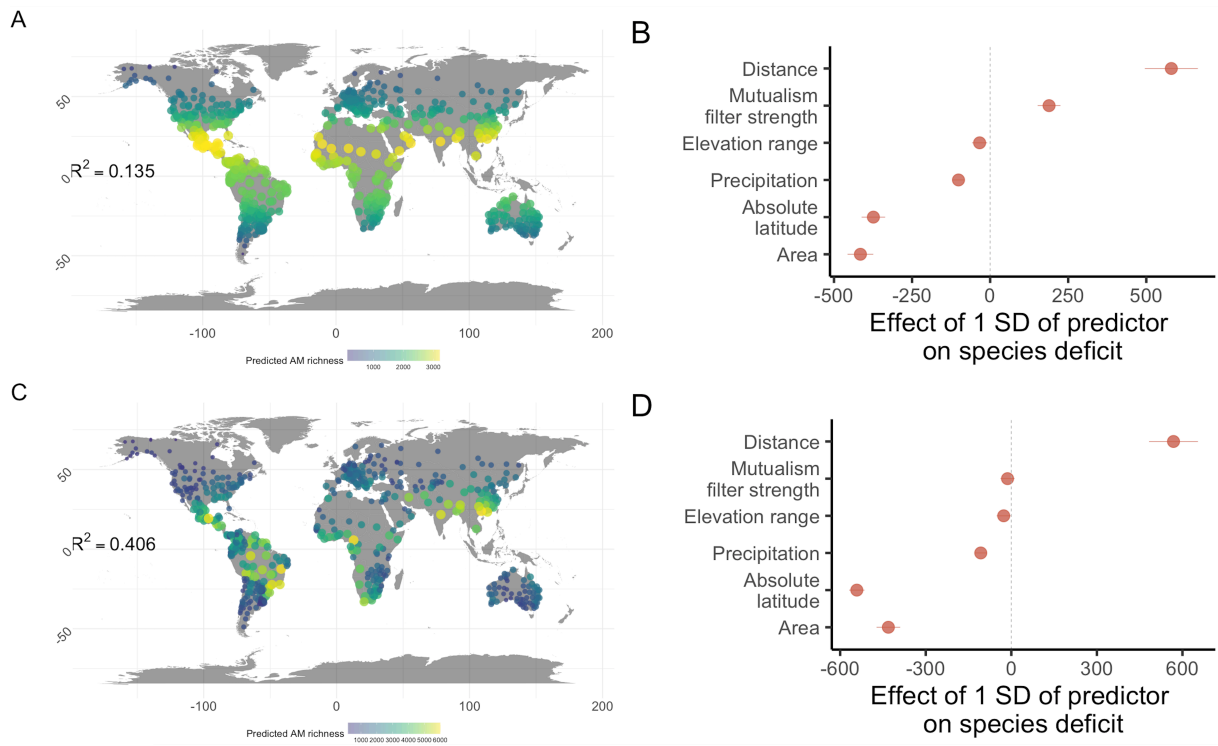

**Figure 1:** Predicted values and corresponding effect estimates for a multiple regression using the original and the refined definition of mutualism strength as predictor. Maps **A** and **C** show for one of the three measures of mutualism (arbuscular mycorrhizal (AM) fungi diversity) how predictions and performance changes between a GAM used by Delavaux et al.[2] with only latitude as a predictor (**A**) ($R^2$ = 0.141) and a random forest model using both longitude and latitude ($R^2$ = 0.406). Results for the other mutualism types were similar (across all three mutualism types, average $R^2$ increased from $R^2$ = 0.149 to $R^2$ = 0.389). Panels **B** and **D** show the results of using these predictions of these two alternative models in a multiple regression. Effects in panel **B** are based on the mutualism strength predicted with latitude only (this corresponds to the original analysis). Effects in panel **D** are based on mutualism strength predicted by our refined model. Note that the effect of mutualism strength on the species deficit is small and not significant.

We believe that this substantial change in the effect estimates for mutualism strength is explained by a second problem in the analysis: Delavaux et al.[2] used a multiple regression with linear effects for all predictors, although a residual analysis shows clear signs of nonlinearities (Fig. S1). This becomes particularly problematic because mutualism strength and latitude are perfectly nonlinearly confounded (Fig. 2A) as the former is predicted only from the latter. Due to this nonlinear correlation of the two variables, it is easily possible that nonlinear effects of latitude induce an apparent effect of mutualism strength. We therefore re-fitted the original multiple regression model with a generalized additive model (GAM)[6] that accounts for such nonlinearities using smooth splines for all variables other than mutualisms filter strength (Fig. 2B, dotted lines). Our results show that area, distance, and absolute latitude influence the species deficit on oceanic islands nonlinearly (Fig. S2), and that after accounting for these nonlinear effects, the effect of mutualism strength (regardless whether we use the original or our revised covariate definition) becomes non-significant (Fig. 2B, red and black dotted lines).

We also show that any misfit in the latitudinal effect, either by treating it linear or by removing it, induces an apparent effect of mutualism strength (Fig. S3). This problem is much reduced by our improved interpolation model for mutualism strength, because using also longitude as a predictor de-correlates mutualism strength from latitude (Fig. 2A; Fig. S3).

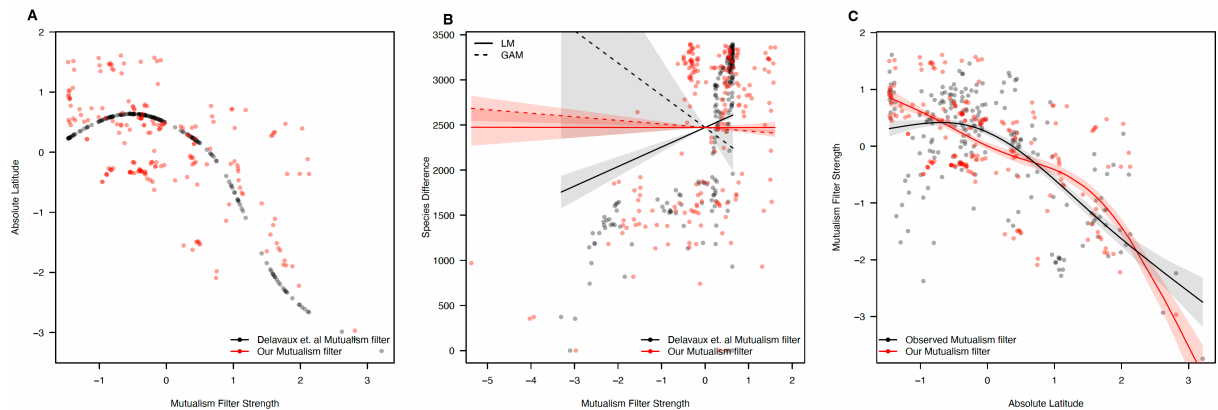

**Figure 2:** Accounting for nonlinear effects removes the reported effect of mutualism strength for both the original and revised definition of this variable, providing additional evidence against a robust effect of mutualists on island species richness. Panel **A**: Due to the methodological choices discussed earlier, the mutualism filter strength used in Delavaux et al.[2] is perfectly non-linear correlated with absolute latitude. Our refined mutualism filter strength variable that is predicted using also longitude shows more variation, which reduces the confounding. Panel **B**: When the original model is extended with splines for all variables except for the mutualism filter strength variable, the effects of mutualism filter strength variables is non-significant, regardless of whether we use the original and the refined predictions for mutualists (Panel **B**, dotted lines). The figure also highlights that the black solid line, corresponding to the original effect found by Delavaux et al.[2], is the only analysis choice that results in a significant effect, whereas either accounting for nonlinearities or using a better predictor of mutualisms results in n.s. effects. Panel **C** shows that there is no difference between the predicted mutualism filter strength on the mainland and the observed mutualism (filter) strength on islands (which we calculated based on the mutualism ratios observed on the islands, rather than predicted based on the mutualism ratios from the mainland), as one would expect if mutualists really had a lower chance of establishment on these islands.

Based on these results, we do not see convincing statistical evidence for a robust effect of the proportion of mainland mutualists species on the species deficit of oceanic islands. We nevertheless asked ourselves if there are secondary patterns that could further help to decide if such an effect exists or not. If it was true that mutualists would establish on oceanic islands at a lower rate because of their missing interaction partners, we would expect the relative proportion of mutualist species on these islands to drop in comparison to the proportion at the mainland. We did not find such an effect (Fig. 2 C): mutualists appear at the same proportion on oceanic islands as they are found on the corresponding mainland. This pattern seems incompatible with the idea that there is a strong establishment disadvantage for mutualists on oceanic islands.

In conclusion, we believe that the relatively strong effect of mutualism rates on species richness in oceanic islands reported by Delavaux et al.[2] may have arisen through the combination of using a poor predictor of mainland mutualism rates that, by design, was nonlinearly confounded with latitude, and a lack of accounting for nonlinearities in the main analysis. After addressing these issues, we do not find convincing statistical evidence for the reported effect (Fig. 1D; Fig. 2B; Fig. S3). The lack of a systematic differences between mutualism strength on islands and the corresponding mainland locations (Fig. 2C) is adding to our doubts if a strong effect of mutualists on island species richness truly exists.

## Code availability

Data and code of our analysis can be found in https://github.com/MaximilianPi/IslandLDG. A nearly complete open dataset to run the analysis is included in the repository. For the result presented here, we used the same restricted data of the original paper, which needs to be requested as described in Delavaux et al.[2]

## Author contributions

MP and FH jointly conceived and designed the study. MP analyzed the data. Both authors contributed equally to the interpretation of the results, as well as the writing and preparation of the manuscript.

## Competing interests

The authors declare no competing interests.

# Supplementary Information S1

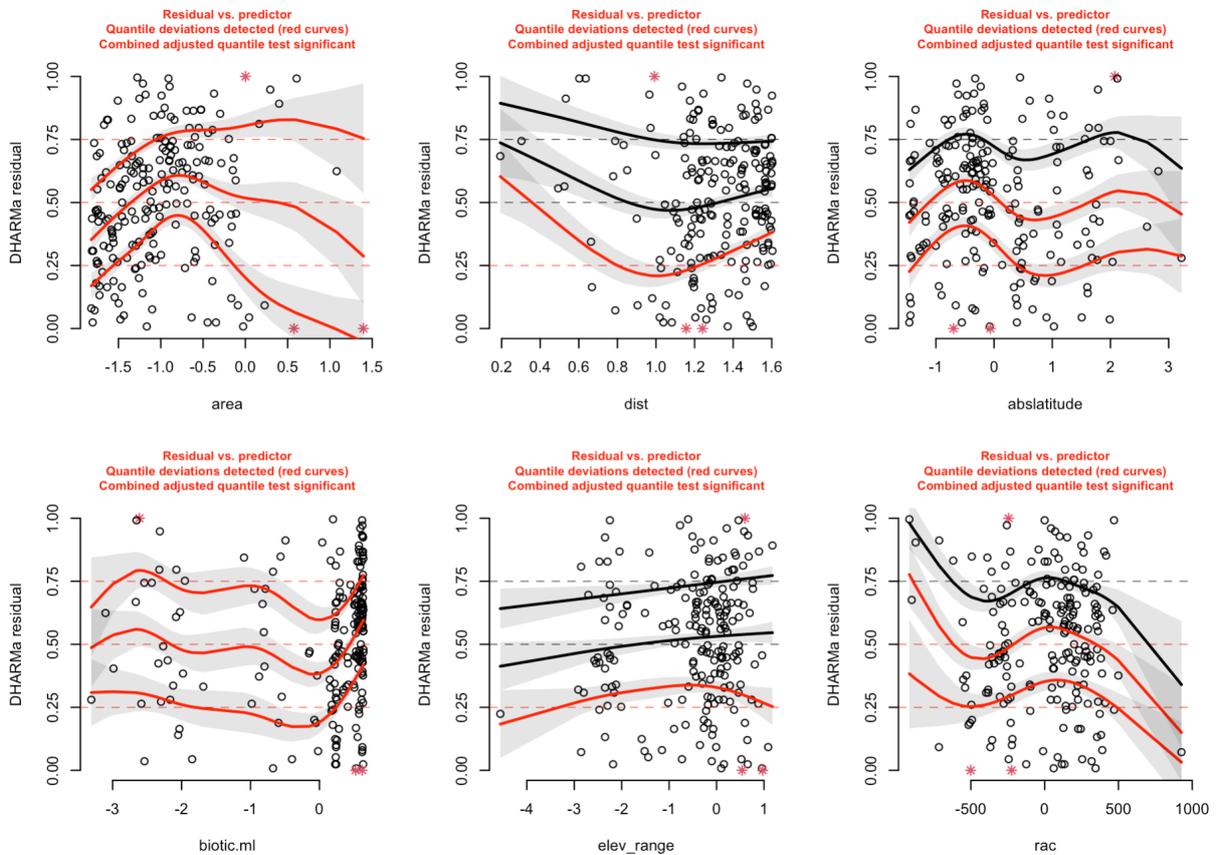

**Figure S1:** Residuals of the model by Delavaux et al.[1]. Most predictors show residuals patterns, indicating nonlinearities that could potentially lead to biased estimates in the case that predictors are nonlinearly correlated. Residuals plots were created using the DHARMa package[2]

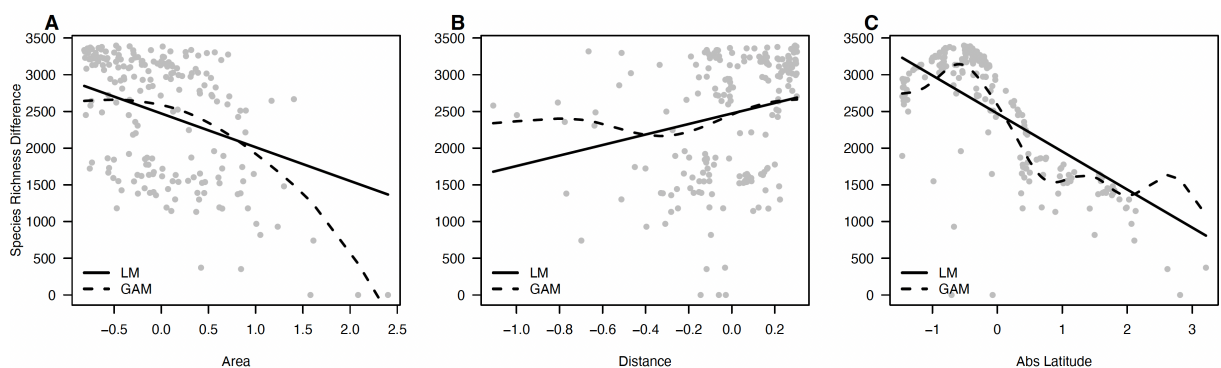

**Figure S2:** Linear and non-linear fits of Area, Distance, and Absolute Latitude based on the simple linear regression used in Delavaux et al.[1] and on our generalized additive model.

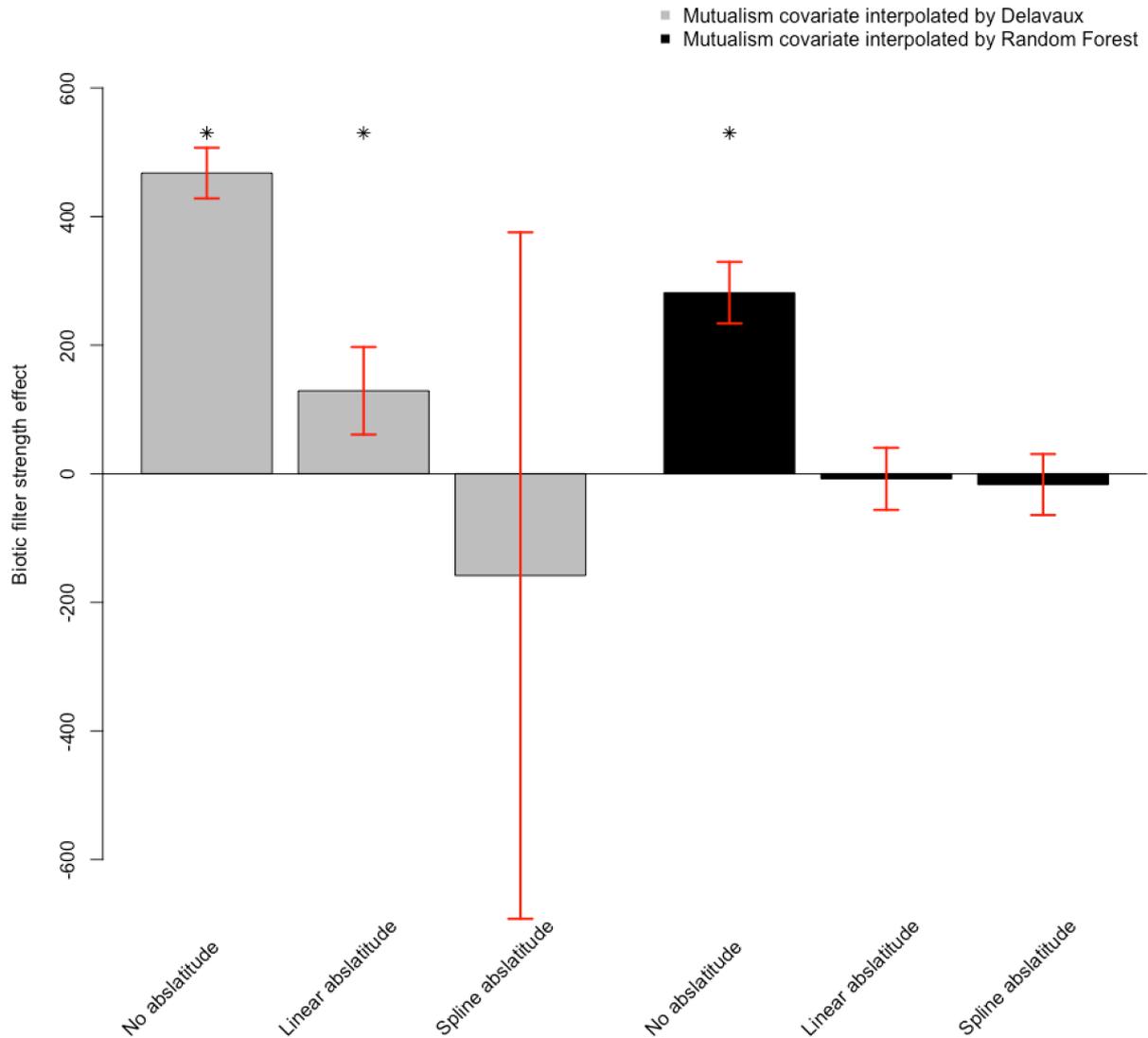

**Figure S3:** The effect of the modelling choice of the absolute latitude variable on the effect estimate of the biotic filter strength variable. Three different ways of modelling absolute latitude were tested: no absolute latitude variable, absolute latitude as a linear effect and absolute latitude as a spline. The biotic filter strength variable was always set as a linear effect. The other variables (distance, area, elevation range, precipitation, and a spatial variable to correct for autocorrelation were set as splines). We extracted the effect estimates (bars) and their confidence intervals (red error bars) for the biotic filter strength variable interpolated by Delavaux et al.[1] (grey) and biotic filter strength variable interpolated by random forest (black). Stars mark significant effects.